\documentclass[RNAAS,linenumbers]{aastex631}

\newcommand\gdrtwo{{\it Gaia}~DR2}

\newcommand{\G}{\textit{Gaia}}
\newcommand{\msun}{${\mathrm{M}}_{\odot}$}
\newcommand{\teff}{\ensuremath{T_{\mathrm{eff}}}}
\newcommand{\logg}{\ensuremath{\log g}}

\shorttitle{DES J2147-4035}
\shortauthors{Apps et al.}
\graphicspath{{./}{figures/}}

\begin{document}

\title{Discovery of a uniquely cool and compact source at 28 parsecs from the Sun}

\author{Kevin Apps}
\affiliation{Independent scholar, United Kingdom, kevinapps73@gmail.com}
\author[0000-0002-4424-4766]{R. L. Smart}
\author{Roberto Silvotti}
\affiliation{
INAF-Osservatorio Astrofisico di Torino,
strada dell'Osservatorio 20,
10025 Pino Torinese, Italy\\
}





\begin{abstract}

We present the discovery of what appears to be both a uniquely cool and old
white dwarf within 30 parsecs of the sun.  DES
J214756.46-403529.3 is detected in four separate surveys, 50 degrees
away from the Galactic Plane.  The combination of its very low
luminosity and spectral energy distribution suggests an object unlike
any other astrophysical body currently known.  Among 8,000 of the
nearest single objects in the immediate solar neighbourhood, it
appears completely isolated in multiple colour-magnitude diagrams.
The data seem compatible with an extremely old and cool white dwarf with a helium 
dominated atmosphere and a mass around 0.7-0.8\,\msun\ but spectroscopic follow-up is 
required to confirm its nature.

\end{abstract}


\section{DES J214756.46-403529.3 properties}
Using an updated and vetted database of objects within 30\,pc we have
identified a source that stands alone in multiple colour-magnitude
diagrams (CMDs). DES J214756.46-403529.3 (DES J2147-4035 from here on) is a
source that has been fully astrometrically parameterised with
consistent parameters in both the \gdrtwo~ and EDR3 releases
\citep[][SourceID 6584418167391671808]{2018A&A...616A...1G, 2021A&A...650C...3G}.  It was found to have
a reliable astrometric solution when tested for inclusion in the \G\
Catalogue of Nearby Stars \citep{2021A&A...649A...6G}. 
Located at 50$^{\circ}$ from the
Galactic Plane, its nearest \G\ EDR3 neighbour is 31” away and there
is nothing brighter than \G\ G = 16 within 100”. It has a good astrometric solution with a parallax error compatible  with its magnitude (35.79$\pm$0.49\,mas); a large number of good observations (577); a low RUWE (1.06) and zero image parameters
determination windows with more than one peak. All this evidence suggests this is a single object with a reliable astrometric solution and no detectable perturbation, either from a
bound source or a background object.  

DES J2147-4035 is visible in all three blue, red and near infrared
Palomar sky survey images with positions consistent with its \G\ EDR3
proper motion.  There is no other object visible at its \G\
EDR3 position in any of the large sky surveys that are available.  The source is detected in all bands
of the Dark Energy Survey DR1 \cite[][g=21.40$\pm$0.02,
  r=19.86$\pm$0.01, i=19.30$\pm$0.01, z=19.11$\pm$0.01,
  Y=18.90$\pm$0.03]{2018ApJS..239...18A}, in the VISTA Hemisphere
Survey DR5 \cite[][J=17.58$\pm$0.02,
  K=17.63$\pm$0.09]{2021yCat.2367....0M} and in the CatWISE survey
\cite[][W1=16.97$\pm$0.05, W2=16.72$\pm$0.13]{2021ApJS..253....8M}.

What makes this object interesting is its
isolated position in a number of CMDs of single objects within 30 parsecs of the Sun. In the figure we reproduce a CMD of all the  \G\
Catalogue of Nearby Stars objects with a \texttt{parallax} $ > 33.333$\,mas.
The figure shows the classical vertical main sequence and the diagonal
white dwarf (WD) cooling sequence on the left. The WD sequence upper edge
is very sharply defined once double degenerates are removed, since the minimum WD mass from single star evolution is
around 0.45\,\msun\ and each effective temperature, or in this
case colour index, corresponds to a reasonably fixed luminosity.  On the right,
the low mass main sequence starts at about M3V at M$_G$=10, extending
down to sub-stellar late L dwarfs at M$_G$=20.

\begin{figure*}
\centering
\includegraphics[width=16cm,angle=0]{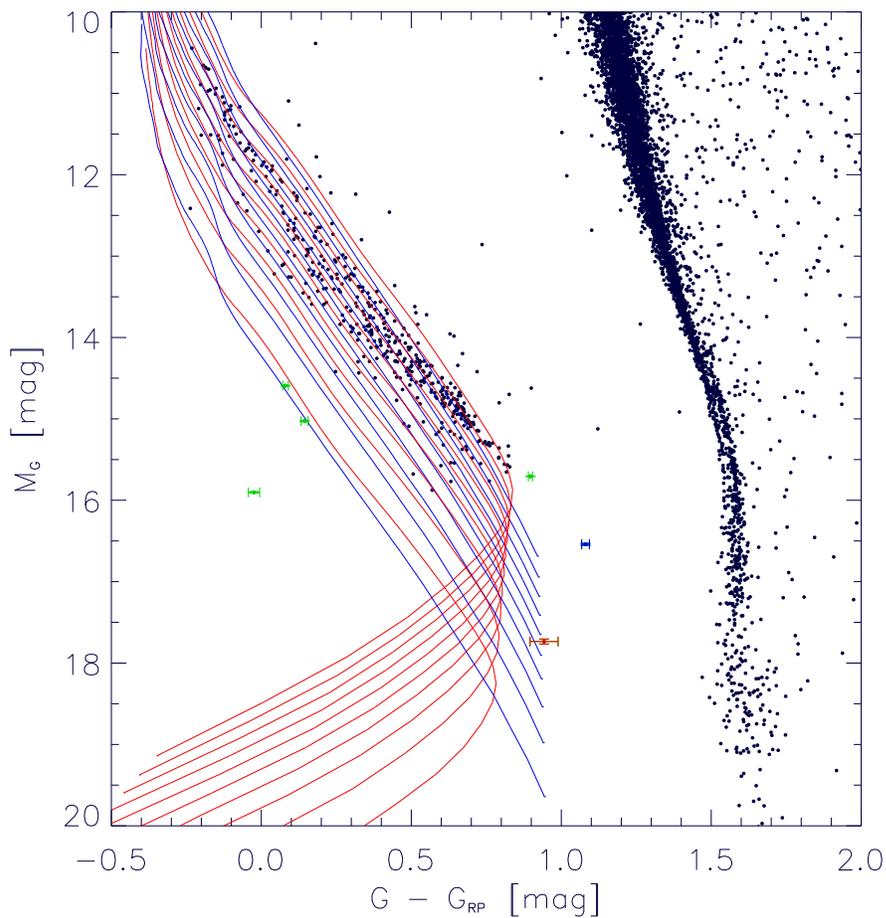}
\caption{Colour-magnitude diagram showing objects within 30\,pc from the Sun. Superimposed are the Montreal cooling tracks for white dwarfs with masses between 0.4 and 1.3\,\msun\ in steps of 0.1\,\msun\ (red is pure H atmosphere or DA WDs, blue is pure He atmosphere or DB WDs). DES J2147-4035 is plotted in red, the green points are four well-studied white dwarfs from left-to-right: LHS~1402, LHS~4033, NLTT~3499, LHS~2229 and the blue point is  EDR3~5616323124310350976, all discussed in the text.}
\label{fig:my_label}
\end{figure*}

We highlight in red the newly identified source in question - DES J2147-4035
and in green four well-studied white dwarfs that also appear somewhat isolated in this
and other CMD plots. From left to right these are LHS 1402, an extremely cool and low
luminosity DC white dwarf that is very blue in optical colours due to strong collisional induced absorption and may have a helium dominated
atmosphere \citep{2005ApJ...625..838B}.
LHS~4033 and NLTT~3499,
the most massive single white dwarfs known out
to 30 parsecs, with a mass of about 1.3\,\msun\, causing their unusually 
small radii and so intrinsically low luminosity  \citep{2021MNRAS.503.5397K}.  
LHS~2229, although seemingly defining the end 
of the white dwarf cooling sequence for low-mass white dwarfs, actually has a larger mass 
(\logg=8.17 and \teff$\simeq$4500 K, \citealt{2012ApJS..199...29G}) and a highly unusual spectral 
energy distribution caused by strong shifted C2 Swan bands that affect the colours. 
Finally, EDR3 5616323124310350976 is an object that we are uncertain of its nature.  It is not visible in the 2MASS survey so it is difficult to be a WD+M dwarf binary inside 30pc, its PanSTARRs colours do not suggest a WD but the proper motion  is visible in comparison of the various surveys. There is a fainter object at \~2" that was not included in the \G\ releases., but the quality flags in the EDR3 do not suggest a corrupted astrometric solution. The \G\ photometry maybe corrupted by the nearby fainter object. This is an object that warrants further spectroscopic followup. 

We can see that DES J2147-4035 sits completely detached from other
objects in the 30\,pc volume. In all CMDs DES J2147-4035 is redder and therefore cooler than the end of the local WD cooling sequence. Its detection by CatWISE2020 in the 4.6 micron W2 band
gives an absolute magnitude of MW2 = 14.49$\pm$0.13.  This is typical
of objects on the T/Y boundary \citep{2021ApJS..253....7K}.
This rules out any object earlier than a Y dwarf being an unresolved
contaminant of this object that could possibly drag its position
towards the red in any CMD.

\section{An extremely old white dwarf\,?}
Without invoking something truly exotic in an astrophysical sense, or having a totally erroneous parallax without any companion EDR3 flags to indicate this, it appears from the available data that DES J2147-4035 has to be an 
extremely cool and old white dwarf.
When we compare its position in the figure with the Montreal cooling tracks 
(https://www.astro.umontreal.ca/~bergeron/CoolingModels,
\citealt{2020ApJ...901...93B} and references therein) we see that it is compatible with an extremely
old white dwarf with a helium-rich atmosphere,
with a mass between 0.7 and 0.8\,\msun\
(when we use G$_{BP}$--G$_{RP}$ instead of
G--G$_{RP}$ in the CMD diagram the best fit is obtained for M$\approx$0.7 \msun). Following these models its cooling age should be around 10\,Gyrs, suggesting a halo or thick disk origin. However, its transverse velocity based purely on \G\ EDR3 data of just 19\,km/s seems in contradiction with this old age. Spectroscopic follow up of DES J2147-4035 is required to confirm the true nature of this faint (G$_{BP}$=20.95$\pm$0.09, G=19.97$\pm$0.01, G$_{RP}$=19.02$\pm$0.05) interesting object.


\bibliography{refs}{}
\bibliographystyle{aasjournal}



\end{document}